# Is it possible to improve existing sample-based algorithm to compute the total sensitivity index?

Samuele Lo Piano[a#], Federico Ferretti[b], Arnald Puy[c], Daniel Albrecht[b],
Stefano Tarantola[d] and Andrea Saltelli[a,c]

[a] Open Evidence Research,
Universitat Oberta de Catalunya (UOC), Barcelona

[b] European Commission, Joint Research Centre, Modelling, Indicators and Impact Evaluation Unit,
Ispra (Italy)

[c] Centre for the Study of the Sciences and the Humanities (SVT),
University of Bergen (UIB, Norway),

[d] European Commission, Joint Research Centre, 'Energy Security, Distribution & Markets' Unit,
Ispra (Italy)

[#] Corresponding Author

**Abstract**

Variance-based sensitivity indices have established themselves as a reference among practitioners of sensitivity analysis of model output. It is not unusual to consider a variance-based sensitivity analysis as informative if it produces at least the first order sensitivity indices $S_j$ and the so-called total-effect sensitivity indices $T_j$ for all the uncertain factors of the mathematical model under analysis.

Computational economy is critical in sensitivity analysis. It depends mostly upon the number of model evaluations needed to obtain stable values of the estimates. While efficient estimation procedures independent from the number of factors under analysis are available for the first order indices, this is less the case for the total sensitivity indices.

When estimating $T_j$, one can either use a sample-based approach, whose computational cost depends on the number of factors, or approaches based on meta-modelling/emulators, e.g. based on Gaussian processes. The present work focuses on sample-based estimation procedures for $T_j$ and tries different avenues to achieve an algorithmic improvement over the designs proposed in the existing best practices. We conclude that some proposed sample-based improvements found in the literature do not work as claimed, and that improving on the existing best practice is indeed fraught with difficulties. We motivate our conclusions introducing the concepts of explorativity and efficiency of the design.

# 1. Introduction

Sensitivity analysis of mathematical models aims 'to apportion the output uncertainty to the uncertainty in the input parameters' (Saltelli and Sobol', 1995). Uses of sensitivity analysis are found in quality assurance, model calibration, model validation, uncertainty reduction, model simplification and many others.

In the last three decades sensitivity analysis (SA) has made steps to establish itself as a discipline on its own, with a community of practitioners gathering around the SAMO (Sensitivity Analysis of Modelling Output) series of international conferences since 1995. Special issues have been devoted to SA (Tarantola and Saint-Geours, 2015, Ginsbourger et al., 2015, Borgonovo and Tarantola, 2012, Turányi, 2008, Helton et al., 2006, Tarantola and Saltelli, 2003), mostly in relation to the SAMO events. Available textbooks for sensitivity analysis include Cacuci, 2003; Saltelli et al., 2000, 2004, 2008; de Rocquigny et al., 2008, Fang et al., 2006. Sensitivity analysis is acknowledged as a useful practice in model development and application. Its use in regulatory setting – e.g. in impact assessment studies – is prescribed in guidelines both in Europe and the United States (European Commission, 2015, p. 390-393; Office for the Management and Budget, 2006, p. 17-18; Environmental Protection Agency, 2009, p.26). Sensitivity analysis is also an ingredient of sensitivity auditing (Saltelli et al., 2013, Saltelli and Funtowicz, 2014), a procedure to investigate the relevance and plausibility of model-based inference as an input to policy (European Commission, 2015, p. 393).

Tools such as sensitivity analysis and sensitivity auditing are particularly needed at this junction in time when the accuracy, relevance and plausibility of statistical and mathematical models used in support to policy is often the subject of controversy (Jakeman et al., 2006; Pilkey and Pilkey-Jarvis, 2007; Saltelli and Funtowicz, 2017, Saltelli and Giampietro, 2017; Padilla et al., 2018). As highlighted elsewhere (Saltelli and Annoni, 2010), part of the problem in the validation of models is that the quality of the accompanying sensitivity analyses is often wanting. Most sensitivity analysis applications seen in the literature still favour the use of a method known as OAT, where the sensitivity of a factor is gauged by moving One-factor-At-a-Time. OAT is still the most largely used technique in SA according to a recent analysis (Ferretti et al., 2016). Thus, sensitivity analysis is run in a perfunctory fashion, failing to be an effective tool to test the model robustness. While different methods exists for sensitivity analysis (see recent reviews in Saltelli et al., 2012, Neumann, 2012, Norton, 2015, Wei et al., 2015, Becker and Saltelli, 2016, and Iooss, 2016, Pianosi et al., 2016), the definition of SA given above - based on decomposing the variance of the output/prediction - is in fact partial to a class of sensitivity analysis methods known as 'variance-based'. These methods are considered as a reference among practitioners. To make an example, when a new method for sensitivity analysis is introduced, its behaviour is investigated by comparing its predictions to those from the variance-based measures, (see e.g. (Mara et al., 2017). At present the most widely used variance-based measures are the Sobol' indices (Sobol', 1990-1993), and in particular Sobol' first order sensitivity measure $S_j$, together with the so-called total sensitivity indices $T_j$, introduced by Homma and Saltelli (1996). In the following we take the suggestion from Glen and Isaacs (2012) to adopt for simplicity the symbol $T_j$, rather than $S_{Tj}$ or $S_j^T$ for the total sensitivity indices, although the latter is also commonly found in the literature.

As the present paper is devoted to the computation of $T_j$, we will immediately focus on Sobol' indices, offering in the following a parsimonious description of how $S_j$ and $T_j$ are defined and computed for the case of independent input factors in order to move speedily to the proposed algorithms. The present work continues an investigation of efficient estimation procedures for $T_j$ (Saltelli, 2002; Saltelli et al., 2010) (outlined in subsections 1.3 and 1.4), looks at the more recent works of Glen and Isaacs (2012) and Lamboni (2018) (subsections 3.1 and 3.2), and proposes new patterns towards computational improvement (subsection 3.3).

## 1.1 Variance-based Sensitivity Measures

For a scalar model output $Y = f(X_1, X_2, \ldots, X_k)$, where $X_1$ to $X_k$ are $k$ uncertain factors the first order sensitivity index $S_j$ can be written as

$$S_j = \frac{V_{X_j}\left(E_{X_{\sim j}}(Y|X_j)\right)}{V(Y)} \qquad (1)$$

where we assume, without loss of generality, factors to be uniformly distributed over the $k$-dimensional unit hypercube $\Omega$.

The inner mean in (1) is taken over all-factors-but-$X_j$, (written as $\mathbf{X}_{\sim j}$), while the outer variance is taken over factor $X_j$. $V(Y)$ is the unconditional variance of the output variable $Y$.

A short recap of this measure should mention without proof (for which see e.g. Saltelli et al., 2008) that:

- An efficient way to estimate $S_j$ is to obtain a curve corresponding to $E_{\mathbf{X}_{\sim j}}(Y|X_j)$ by smoothing or interpolation of the scatterplot of $Y$ versus the sorted values of variable $X_j$ and then compute the variance of this curve over $X_j$, see Figure 1.
- $S_j$ coincides with the squared Pearson correlation ratio, customarily indicated as $\eta^2$.
- $S_j$ reduces to the squared value of the standardized regression coefficient, i.e. to $\beta^2$, when the relationship between $Y$ and $X_j$ is linear.
- $S_j$ is a first order term in the variance decomposition of $Y$ (valid when the input factors are independent) which includes terms up to the order $k$, i.e.

$$1 = \sum_{j=1}^{k} S_j + \sum_{j<l} S_{jl} + \cdots + S_{12\ldots k} \qquad (2)$$

- Generic terms $S_{j_1 j_2 \ldots j_n}$ higher than the first order ones $S_j$ are used sparingly in applications due to their multiplicity: a model with $k = 3$ has just three second order terms, but one with $k = 10$ has as many as 45 second order terms. The total number of terms in (2) is $2^k - 1$.
- The meaning of $S_j$ in plain English is 'the fractional reduction in the variance of $Y$ which would be obtained on average if $X_j$ could be fixed'. This derives from another useful relationship:

$$V_{X_j}\left(E_{\mathbf{X}_{\sim j}}(Y|X_j)\right) + E_{X_j}\left(V_{\mathbf{X}_{\sim j}}(Y|X_j)\right) = V(Y) \qquad (3)$$

The second term in (3) is in fact the average of all partial variances obtained fixing $X_j$ to a value over its uncertainty range – thus the first term in (3) results to be the average reduction. Note that while $V_{\mathbf{X}_{\sim j}}(Y|X_j)$ can be greater than $V(Y)$, $E_{X_j}\left(V_{\mathbf{X}_{\sim j}}(Y|X_j)\right)$ is always smaller than $V(Y)$ because of (3).

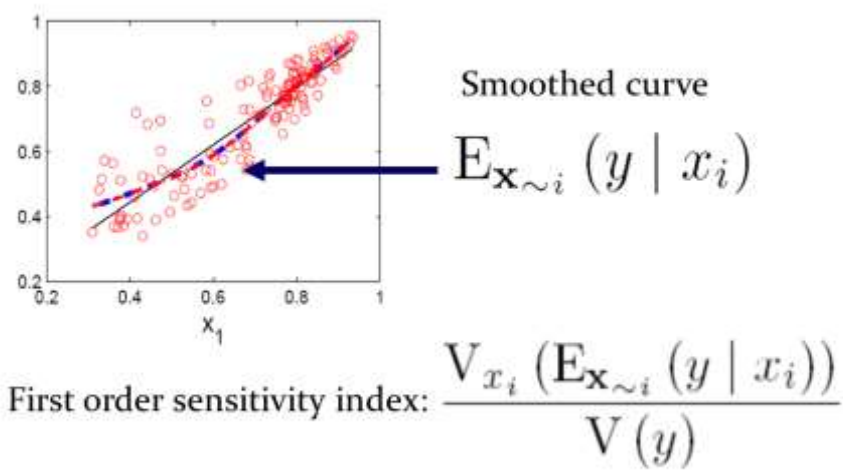

*Figure 1 Sensitivity measures and their relationships in an hypothetical Y(X₁) chart. The dashed line represents the local mean of the points in the scatterplot, while the straight line corresponds to Pearson's correlation ratio $\eta^2$.*

The total order sensitivity coefficients $T_j$ can be written as:

$$T_j = \frac{E_{\boldsymbol{X}_{\sim j}}\left(V_{X_j}(Y|\boldsymbol{X}_{\sim j})\right)}{V(Y)} \tag{4}$$

Worth recalling for this measure is that:

- Unlike the case of $S_j$ the smoothing/interpolation of the inner variance is precluded by the impossibility to sort by $\boldsymbol{X}_{\sim j}$ other than using emulators which are not the subject of the present paper.
- Its meaning in English – explicitly descending from Equation (4) - is 'the fraction of variance that would remain on average if $X_j$ would be left free to vary over its uncertainty range while all other factors were fixed'.
- Applying again (3) one gets

$$E_{\boldsymbol{X}_{\sim j}}\left(V_{X_j}(Y|\boldsymbol{X}_{\sim j})\right) + V_{\boldsymbol{X}_{\sim j}}\left(E_{X_j}(Y|\boldsymbol{X}_{\sim j})\right) = V(Y) \tag{5}$$

Noting that the second term in (5) is the first order effect on all-but-$X_j$, one derives that the first term in (5), i.e. the numerator in the equation (4), is the total variance of all terms in decomposition (2) which do include factor $X_j$. E.g. for a model with just three factors one can write

$$1 = S_1 + S_2 + S_3 + S_{12} + S_{13} + S_{23} + S_{123}$$

and

$$T_1 = S_1 + S_{12} + S_{13} + S_{123}$$

- Hence a parsimonious description of the model $= f(X_1, X_2, \ldots, X_k)$, which nevertheless tells us which factor behaves additively $(S_j = T_j)$ and which does not $(S_j < T_j)$, can be obtained by computing all $k$ terms $S_j$ and all $k$ terms $T_j$. For an additive model it holds that $S_j = T_j \ \forall \ j$ and $\sum_{j=1}^{k} S_j = 1$.

It is acknowledged by practitioners that one of the main problems with using the couple $S_j, T_j$ is the computational cost, which can be a problem when $Y = f(X_1, X_2, \ldots, X_k)$ is computationally time consuming. This could be the case for a mathematical model involving large system of differential equations, for a laborious optimization program, for natural system simulators involving spatially distributed grid points, and so on. This difficulty is especially relevant for $T_j$, as we go on to discuss in the next section.

| | *Table 1 - Legend* |
|---|---|
| $\boldsymbol{A}, \boldsymbol{B}, \boldsymbol{C}, \ldots$ | Sample matrices |
| $\boldsymbol{A}_{\boldsymbol{B}}^{(j)}$ | Sample matrix where all columns are from $\boldsymbol{A}$ but for column $j$ which is from $\boldsymbol{B}$; likewise for other combinations of sample matrices $\boldsymbol{C}, \boldsymbol{D}, \boldsymbol{E}$ and so forth. |
| $\boldsymbol{a}_i, \boldsymbol{b}_i, \boldsymbol{c}_i, \ldots$ | $i^{th}$ row of respectively matrices $\boldsymbol{A}, \boldsymbol{B}, \boldsymbol{C}$. |
| $\boldsymbol{a}_{bi}^{(j)}$ | $i^{th}$ row of matrix $\boldsymbol{A}_{\boldsymbol{B}}^{(j)}$ |
| $e$ | Economy of a given design, defined as the number of elementary effects per run |
| $E_{X_j}(\cdot), V_{X_j}(\cdot)$ | Expected value and variance of argument $(\cdot)$ taken over factor $X_i$ |

| $E_{X_{\sim j}}(\cdot), V_{X_{\sim j}}(\cdot)$ | Expected value and variance of argument $(\cdot)$ taken over factor all factors but $X_i$ |
|---|---|
| $h$ | Generic sample matrix |
| $i$ | Running index for the rows of a sample matrix $i = 1,2,\ldots N$ |
| $j$ | Running index for factor number $j = 1,2,\ldots k$ |
| $k$ | Number of factors |
| $l$ | Running index over factor number $j$; $l = 1,2,\ldots j$ |
| $m, q$ | Running indices on the pool of sample-matrices |
| $n$ | Number of sample matrices |
| $N$ | Column dimension (length) of a single sample matrix |
| $p$ | Running index on the block on which the algorithm is executed (each block has column length $N = 2^p$) |
| $r$ | Running index for the repetition $r = 1,2,\ldots 50$ |
| $S_j$ | First order effect sensitivity index |
| $T_j$ | Total effect sensitivity index |

## 1.2 Sample-based Estimation procedures

The evaluation of sensitivity indices is often based on Monte Carlo integration. Monte Carlo based procedures for the estimation of $S_j$ have been proposed by Sobol' (1990-1993), McKay (1995), Saltelli (2002), Ratto et al. (2007), Sobol' et al. (2007), Mara and Rakoto-Joseph (2008), Lilburne and Tarantola (2009), Saltelli et al., 2010, Glen and Isaacs (2012), Owen (2013), Plischke et al. (2013) and Janon et al. (2014). Some particularly efficient algorithms for the estimation of $S_j$ belong to the class of spectral methods, which may be preferred in case the model has some regularity (Prieur and Tarantola, 2016). Among these we can mention random balance designs (Tarantola et al., 2006), and an "Effective Algorithm to compute Sensitivity Indices – EASI" (Plischke, 2010), both requiring a total number of model evaluations that is independent on the number of factors. Although two spectral methods were devised by Saltelli et al. (1999) and Mara (2009), the estimation of $T_j$ mostly relies on Monte Carlo-based approaches. See, for instance, Homma and Saltelli (1996), Saltelli (2002), Lilburne and Tarantola (2009), Saltelli et al. (2010), Buzzard and Xiu (2011), Glen and Isaacs (2012) and Kucherenko et al. (2015).

In this paper we focus on the sole $T_j$ measure and on those estimations based on actual evaluation of the function at the sampled points, e.g. without recourse to meta-modelling approaches. We move from the recipe given in Saltelli et al., 2010, which is in turn derived from Saltelli, 2002. A quick recap of the main ingredients of this recipe is as follows:

- The computation is based on a quasi Monte Carlo (QMC) method and makes use of quasi-random (QR) points of Sobol' $LP_\tau$ sequences (Sobol', 1967, 1976). Quasi random sequences possess desirable uniformity property over the unit hypercube $\Omega$, a $k$-dimensional cube of side equal to one. QR numbers are not random. They are designed to optimally fill the unit hypercube in the sense of avoiding inhomogeneity (clustering) of points.
- A useful concept in this respect is that of discrepancy. Given a set of $M$ points inside $\Omega$, their discrepancy is the maximum deviation, over all possible parallelepipeds drawn within the hypercube $\Omega$, of the theoretical density ($M$ times the volume of the parallelepiped) versus the actual density. Sobol' $LP_\tau$ sequences are designed to be 'low discrepancy' and perform well in existing QR method inter-comparisons (e.g. Bratley and Fox, 2008).
- We use here two different sequence generators for Sobol' points: the Algorithm 659 (Bratley and Fox, 1988) and SobolSeq16384, distributed by BRODA Ltd. (2015) and based on Sobol'

- et al. (2011) for three different software implementations of our experiment (Python and Rbased on the former, whereas Matlab®- on the latter).
- The method uses sample matrices of row dimension $k$, the number of factors, and a column dimension $N$, which - as we shall see - will be influent in determining the total cost of the estimation procedure. We start from two $N \times k$ matrices $\boldsymbol{A}$ and $\boldsymbol{B}$. Those matrices are constructed from an LP$_\tau$ sequence of column dimension $N$ and row dimension $2k$, using the leftmost $k$ columns for $\boldsymbol{A}$ and the remaining columns for $\boldsymbol{B}$.
- A relevant characteristic of LP$_\tau$ sequences is that the uniformity properties deteriorate moving from left to right along the row of the sequence. This means that, for any given $N$, one would expect that matrix $\boldsymbol{A}$ has lower discrepancy than $\boldsymbol{B}$.
- The estimation of $T_j$ requires points in $\Omega$ which are separated by what is called a 'step in the $X_j$ direction', or in other words two points in $\Omega$ which only differ in the value of factor $X_j$. Note that $S_j$ requires instead steps in the non-$X_j$ direction, e.g. couples of points where all factors but $X_j$ have differing values. As discussed elsewhere (Campolongo et al., 2011), the estimation procedure for $T_j$ resembles that for the method of Morris: $T_j$ is to be preferred to Morris even for factors' screening purposes as $T_j$ estimation requires less modelling assumptions than Morris.
- Given the two $N \times k$ matrices $\boldsymbol{A}$ and $\boldsymbol{B}$, we build an additional set of $k$ matrices which we indicate as $\boldsymbol{A}_{\boldsymbol{B}}^{(j)}$, where column $j$ comes from matrix $\boldsymbol{B}$ and all other $k-1$ columns come from matrix $\boldsymbol{A}$. We indicate as $\boldsymbol{a}_i$ the $i^{th}$ row of $\boldsymbol{A}$ and likewise with $\boldsymbol{a}_{bi}^{(j)}$ the $i^{th}$ row of $\boldsymbol{A}_{\boldsymbol{B}}^{(j)}$. Thus $\boldsymbol{a}_{bi}^{(j)}$ is the $i^{th}$ row a matrix whose columns come from $\boldsymbol{A}$ but for column $j$ which comes from $\boldsymbol{B}$.
- The model $Y = f(X_1, X_2, \ldots, X_k)$ is run at all $f(\boldsymbol{a}_i)$ and $f\left(\boldsymbol{a}_{bi}^{(j)}\right)$ points, at a cost of $N(k+1)$ model runs, i.e. $N$ times for the $f(\boldsymbol{a}_i)$ points and $Nk$ times for the $f\left(\boldsymbol{a}_{bi}^{(j)}\right)$ points.
- The numerator in equation (4), needed to compute $T_j$, is obtained from the estimator (Jansen, 1999):

$$E_{\boldsymbol{X}_{\sim j}}\left(V_{X_j}(Y|\boldsymbol{X}_{\sim j})\right) = \frac{1}{2N}\sum_{i=1}^{N}\left[f(\boldsymbol{a}_i) - f\left(\boldsymbol{a}_{bi}^{(j)}\right)\right]^2 \qquad (6)$$

- The total variance, the denominator of equation (4), has been estimated using independent runs. These correspond to the subsequent rows of the (independent) matrix(ces) $\boldsymbol{A}, \boldsymbol{B}, \boldsymbol{C}$, etc. depending on the estimator one is assessing.
- Each summand in equation (6) constitutes an elementary effect fungible for the computation of the total sensitivity index $T_j$ for factor $X_j$.

Sobol (2001) noted that formula (6) was originally proposed by Šaltenis et al., 1982 (in Russian), so in the following we shall call it after Šaltenis. Sobol' (2001) also proves (Theorems 3, 4 therein) that Equation (6) has lower variance than a popular competing estimator which we do not discuss here.

Following Glen and Isaacs (2012), we enrich our quest for more efficient design and estimation procedures by trying additional estimators which are based on Sobol' and Levitan (1999), whereby $T_j$ is derived from the expectation value of the Pearson correlation coefficient between the vectors composed of the $N$ elements of $f(\boldsymbol{a}_i)$ and $f\left(\boldsymbol{a}_{bi}^{(j)}\right)$. These formulae are described in a later section together with the enhancement suggested by Glen and Isaacs (2012). We shall also compare the design of Saltelli (2010) against that of Glen and Isaacs (2012) and that of Lamboni (2018). Our experiment uses a broader set of the test functions than that used in these two papers.

In the next section we generalize the computation of the total sensitivity measures to a number $n$ of matrices larger than two in order to propose our sample-based acceleration of the computational procedure. Note that Owen (2013) also proposes the use of three matrices as a more efficient design, but only deals with the estimation of first-order indices.

### 1.3 Exploring the unit cube

As discussed in Saltelli et al., 2010, there is a natural trade-off between the economy of a design – i.e. how many effects we can obtain with a given number of runs, and how well the design fills or explores the input factors space, i.e. in our case the unit hypercube $\Omega$.

The trade-off between economy and exploring the hypercube comes from the fact that the larger $n$ and/or $k$, the more we are re-using coordinates. The $nN$ points corresponding to the $n$ matrices $\boldsymbol{A}$, $\boldsymbol{B}$, $\boldsymbol{C}$,… will all contain original coordinates, while all the hybrid matrices such as $\boldsymbol{A}_{\boldsymbol{B}}^{(i)}$ are re-using coordinates – be it that the points are different, i.e. no point in $\boldsymbol{A}_{\boldsymbol{B}}^{(i)}$ coincides with points of either $\boldsymbol{A}$ or $\boldsymbol{B}$ while e.g. $\boldsymbol{A}$ and $\boldsymbol{A}_{\boldsymbol{B}}^{(i)}$ share $k-1$ columns for a total of $N(k-1)$ coordinates. We shall show in a moment how – for a fixed total number of runs $N_T$ – increasing the number of matrices $n$ decreases the number of non-repeated coordinates.

As discussed in Saltelli et al., (2010) the economy of a design to compute sensitivity indices can be measured in terms of the number of elementary effects obtained per run. From the previous section we know that the design just described yields $N$ elementary effects per factor, i.e. $Nk$ differences $f(\boldsymbol{a}_i) - f\left(\boldsymbol{a}_{bi}^{(j)}\right)$ used in equation (6) at the cost of $N(k+1)$ runs of the model. The economy of the design is thus:

$$e = \frac{Nk}{N(k+1)} = \frac{k}{k+1}, \tag{7}$$

which is less than one. From now on we shall call this design 'asymmetric' for the different roles attributed to the matrices $\boldsymbol{A}$ and $\boldsymbol{B}$ in the design – the coordinates of $\boldsymbol{A}$ are used more than those of $\boldsymbol{B}$.

In the same Saltelli et al. (2010) work an alternative approach was also tried based on the idea of increasing the number of matrices, i.e. instead of using just matrices $\boldsymbol{A}$ and $\boldsymbol{B}$ to take a larger number $n > 2$ of matrices. The idea was that with e.g. $n = 3$ matrices $\boldsymbol{A}$, $\boldsymbol{B}$ and $\boldsymbol{C}$, and using always $\boldsymbol{A}$ as the base sample matrix, one would dispose of $\binom{n}{2} = \binom{3}{2} = 3$ ways of generating elementary effects e.g. using not only couples of function values $f(\boldsymbol{a}_i), f\left(\boldsymbol{a}_{bi}^{(j)}\right)$, but also the couples $f(\boldsymbol{a}_i), f\left(\boldsymbol{a}_{ci}^{(j)}\right)$ and $f\left(\boldsymbol{a}_{bi}^{(j)}\right), f\left(\boldsymbol{a}_{ci}^{(j)}\right)$. All these couples are in fact only one step $x_i$ apart. This design gives $3Nk$ elementary effects at the cost of $N(1 + 2k)$ runs, i.e. an economy of $3k/(1 + 2k)$, greater than one. For a generic $n$ the economy $e$ is given by the total number of effects $E_T$ divided the total number of runs $N_T$. For a generic number $n$ of matrices we have (eq. 8-10):

$$N_T = N(1 + k(n-1)) \tag{8}$$

$$E_T = \binom{n}{2} kN \tag{9}$$

$$e = \frac{E_T}{N_T} = \frac{\binom{n}{2}k}{1 + k(n-1)} = \frac{kn(n-1)}{2(1 + k(n-1))} \tag{10}$$

Trying values of $n > 2$ in Saltelli et al., 2010[1] resulted in poorer convergence with respect to the case $n = 2$ so that the paper confirmed the use of the Šaltenis estimator in conjunction with $n = 2$ as the best available sample-based practice. Contrasting findings have been reported by (Lamboni, 2018), according to whom the optimal number of matrices may be different from two dependent on the function being evaluated.

Glen and Isaacs (2012) still use two matrices, but in a rather different way by dropping the concept of base sample (the role entrusted to matrix $\boldsymbol{A}$), and treating matrices $\boldsymbol{A}$ and $\boldsymbol{B}$ equally (symmetrically). In the present paper we compare the symmetric versus asymmetric approaches. We also try to investigate what happens when the number of matrices is greater than two. We do this using the symmetric design, i.e. treating all matrices equally.

We illustrate this again for the case $n = 3$ before generalizing it. We start by computing all function values for the three matrices (and not just for $\boldsymbol{A}$), i.e. we compute the $3N$ function values $f(\boldsymbol{a}_i), f(\boldsymbol{b}_i), f(\boldsymbol{c}_i)$. Next we compute function values corresponding to all possible mixed matrices e.g. not only $f\left(\boldsymbol{a}_{bi}^{(j)}\right)$ and $f\left(\boldsymbol{a}_{ci}^{(j)}\right)$, but also $f\left(\boldsymbol{b}_{ai}^{(j)}\right), f\left(\boldsymbol{b}_{ci}^{(j)}\right)$ and $f\left(\boldsymbol{c}_{ai}^{(j)}\right), f\left(\boldsymbol{c}_{bi}^{(j)}\right)$ as well, where the usual notation has been kept, i.e. matrix $\boldsymbol{B}_A^{(j)}$ has all columns from $\boldsymbol{B}$ but for column $j$ which is taken from $\boldsymbol{A}$. We have thus generated an additional set of $6Nk$ function values for a total of $3N(1 + 2k)$ runs. How many elementary effects can we compute now? Each of the three matrices $\boldsymbol{A}$, $\boldsymbol{B}$ and $\boldsymbol{C}$ can be used to compute $2Nk$ effects, see the first six rows of Table 2 below. This gives in total $6Nk$ elementary effects. There are additional effects which can be obtained by mixing the hybrid matrices, see the last three rows in Table 2, which give $3Nk$ additional effects.

| | |
|---|---|
| *Table 2 Elementary effects, i.e. couples of function values fungible for the computation of $T_j$, for the case of $n = 3$ matrices* | |
| $f(\boldsymbol{a}_i)$ | $f\left(\boldsymbol{a}_{bi}^{(j)}\right)$ |
| $f(\boldsymbol{a}_i)$ | $f\left(\boldsymbol{a}_{ci}^{(j)}\right)$ |
| $f(\boldsymbol{b}_i)$ | $f\left(\boldsymbol{b}_{ai}^{(j)}\right)$ |
| $f(\boldsymbol{b}_i)$ | $f\left(\boldsymbol{b}_{ci}^{(j)}\right)$ |
| $f(\boldsymbol{c}_i)$ | $f\left(\boldsymbol{c}_{ai}^{(j)}\right)$ |
| $f(\boldsymbol{c}_i)$ | $f\left(\boldsymbol{c}_{bi}^{(j)}\right)$ |
| $f\left(\boldsymbol{a}_{bi}^{(j)}\right)$ | $f\left(\boldsymbol{a}_{ci}^{(j)}\right)$ |
| $f\left(\boldsymbol{b}_{ai}^{(j)}\right)$ | $f\left(\boldsymbol{b}_{ci}^{(j)}\right)$ |
| $f\left(\boldsymbol{c}_{ai}^{(j)}\right)$ | $f\left(\boldsymbol{c}_{bi}^{(j)}\right)$ |

This gives in total $9Nk$ effects for the case $n = 3$ matrices. We have in the end an economy $e = 9Nk / 3N(1 + 2k) = 3k / (1 + 2k)$. How can this be generalized to a design with a generic number of matrices? Given $n$ matrices there are $\binom{n}{2}$ pairwise combinations of these and for each combination $2k$ matrices are produced, i.e. the number of factors times two since for each couple of matrices such as $\boldsymbol{A}$ and $\boldsymbol{B}$ we shall have to consider both matrices $\boldsymbol{B}_{\boldsymbol{A}}^{(j)}$ and $\boldsymbol{A}_{\boldsymbol{B}}^{(j)}$.

---

[1] Equation (8) differs from Equation (22) in Saltelli et al. (2010) because in that work we used $n$ to indicate the number of matrices on top of the base matrix. Replacing $n$ with $n - 1$ in Equation (22) of that paper gives our Equation (8).

Since each matrix is composed of $N$ runs the total number of runs will be $N\left(n + 2k\binom{n}{2}\right) = N(n + kn(n-1)) = nN(1 + k(n-1))$.

With similar considerations one derives that the number of effects will be $Nkn(n-1) + 3Nk\binom{n}{3} = Nkn(n-1) + \frac{1}{2}Nkn(n-1)(n-2) = \frac{1}{2}Nkn^2(n-1)$.

In summary we have

$$N_T = nN(1 + k(n-1))$$

$$E_T = \frac{1}{2}Nkn^2(n-1),$$

and the resulting economy is the one defined in Eq.(10). The value of $e$ tends to $n/2$ for large $n$ or/and large $k$. Note that the same development made for $T_j$ could be replicated for the effects of type $S_j$. The reader can easily verify that for each $T_j$ effect derived above an $S_j$ effect exists as well for the same set of runs. Thus, the economy of the design in terms of both $S_j$ and $T_j$ effects would be double. However, we are not interested here in the estimation of the $S_j$ terms and rest with our definition of economy (10) based purely on $T_j$ effects. As mentioned above, the trade-off between economy and exploration is due to the fact that $e$ grows with $n$ while the opposite is the case for the explorativity of the design – the larger $n$, the larger the number of coordinates which are repeated. The same trade off can be visualized in terms of original versus repeated in a given design, see Appendix 1.

## 1.4 The approach proposed by Glen and Isaacs (2012)

We are now ready to describe the estimators used in Glen and Isaacs (2012) based on computing the Pearson correlation coefficients between vectors. This means that for each of the couples of vectors just described (see e.g. Table 2) instead of applying Šaltenis estimator (6) we compute first the correlation coefficients e.g. for the first entry in Table 2:

$$\rho_j = \frac{1}{(N-1)}\sum_{i=1}^{N}\frac{(f(\boldsymbol{a}_i) - \langle f(\boldsymbol{a}_i)\rangle)\left(f\left(\boldsymbol{a}_{bi}^{(j)}\right) - \langle f\left(\boldsymbol{a}_{bi}^{(j)}\right)\rangle\right)}{\sqrt{V(f(\boldsymbol{a}_i))V\left(f\left(\boldsymbol{a}_{bi}^{(j)}\right)\right)}}, \qquad (11)$$

Where e.g. $\langle f(\boldsymbol{a}_i)\rangle$ is the mean of the $f(\boldsymbol{a}_i)$'s over the $N$ runs and $V(f(\boldsymbol{a}_i))$ their variance. In order to avoid cluttering of the symbols we have not indicated the dependence of $\rho_j$ upon the selected couples of function values $f(\boldsymbol{a}_i)$ and $f\left(\boldsymbol{a}_{bi}^{(j)}\right)$. A mean value $\langle \rho_j \rangle$ of $\rho_j$ is then computed over all possible couples and this enters into the computation of the total sensitivity index as:

$$T_j = 1 - \langle \rho_j \rangle \qquad (12)$$

Additionally, Glen and Isaacs (2012) note that for finite values of $N$ there will be spurious correlations even between supposedly uncorrelated vectors such as $f(\boldsymbol{a}_i)$ and $f(\boldsymbol{b}_i)$ or $f\left(\boldsymbol{a}_{bi}^{(j)}\right)$ and $f\left(\boldsymbol{b}_{ai}^{(j)}\right)$. We say 'supposedly uncorrelated' as no columns are shared between $f(\boldsymbol{a}_i)$ and $f(\boldsymbol{b}_i)$, nor is this the case for $f\left(\boldsymbol{a}_{bi}^{(j)}\right)$ and $f\left(\boldsymbol{b}_{ai}^{(j)}\right)$. These spurious correlations are explicitly computed in Glen and Isaacs (2012) and then used as correction terms in the computation of the sensitivity indices.

The main advantage of the symmetric design proposed Glen and Isaacs (2012) with respect to the asymmetric design of Saltelli et al. (2010) is that in the latter the coordinates of the base sample appear disproportionately with respect to the other coordinates, while this is not the case with the former (whether $n = 2$ or higher). In the next sections we describe the experimental set up and provide our results for the following experiments:

1. Testing the estimator of Glen and Isaacs (2012);
2. Testing the asymmetric versus symmetric design;
3. Identifying optimal values for $n$ as to improve on the existing practices.

## 2 Experimental set up - Test cases

The main Python code used for the computations reported in the present work is available at the following *GitHub* repository: https://github.com/Confareneoclassico/New_estimator_algorithm. A second Matlab® code was used in a separate set of computations limited to one test functions. Finally, a third set of scripts has been coded in *R* to replicate the results demonstrated in subsection 3.3 for one of the tested approaches. These scripts are also available from a *GitHub* repository: https://github.com/arnaldpuy/new-STi-estimator. The agreement of the independent computations coded and run by separate co-authors (SLP, FF and AP) at different points in time and places is taken as an internal validation of the results presented in this paper.

Both Saltelli et al., (2010) and Glen and Isaacs (2012) base their analysis on a single function, namely the G function defined as:

$$G(X_1, X_2, \ldots, X_k, a_1, a_2, \ldots, a_k) = \prod_{j=1}^{k} g_j, \tag{13}$$

with

$$g_j = \frac{|4X_j - 2| + a_j}{1 + a_j} \tag{14}$$

With this test function one can modulate the importance of a factor $X_j$ via its constant coefficient $a_j$, see Table 3.

| $a_j$ | Importance of $X_j$ |
|---|---|
| 0 | Very high |
| 1 | High |
| 9 | Non-important |
| 99 | Non-significant |

*Table 3 – Factors' importance in G as a function of constant parameter $a_j$*

Although this function can be seen as attributed to Davis and Rabinowitz (1984), it is in fact due to Sobol' (Saltelli and Sobol' 1995) and is known among practitioners as Sobol's *G* function. It reduces to a function used in Davis and Rabinowitz (1984) for the particular case of $a_j$=0. The Sobol's *G* function is particularly suited for sensitivity analysis exercises and has in fact been used systematically in the last two decades for this purpose.

A six-dimensional version of the *G* function was used, with coefficients $a_j$ equal to {0 0.5 3 9 99 99} both in Saltelli et al., (2010) and Glen and Isaacs (2012), with decreasing order of importance for the six factors.

In order to test the effectiveness of the estimators with a wider typology of functions, a pool of functions suggested by Kucherenko et al., 2011, equations 15 – 21, has also been tested in the *Python* and *R* implementations.

- $A1: f(X) = \sum_{j=1}^{k}(-1)^i \prod_{l=1}^{j} x_j$ (15)

- $A2: f(X) = \prod_{j=1}^{k} \frac{|4x_j-2|+a_j}{1+a_j}$ (16)

- $B1: f(X) = \prod_{j=1}^{k} \frac{k-x_j}{k-0.5}$ (17)

- $B2: f(X) = (1+\frac{1}{k})^k \prod_{j=1}^{k} \sqrt[k]{x_j}$ (18)

- $B3: f(X) = \prod_{j=1}^{k} \frac{|4x_j-2|+a_j}{1+a_j}$ (19)

- $C1: f(X) = \prod_{j=1}^{k} |4x_j - 2|$ (20)

- $C2: f(X) = 2^k \prod_{j=1}^{k} x_j$ (21)

The analytical values for the sensitivity indices of these functions are available (Saltelli et al., 2010; Kucherenko et al., 2011). Function A2, Equation 16, is the same as Equation 14, where the set of coefficients $a_j$ have been used: {0 0.5 3 9 99 99}.

Following the taxonomy of these authors the functions of the group A are the easiest for sensitivity analysis, as there are few important variables. In class B functions instead, what is low is the number of interaction terms, while in class C, the most difficult to treat for SA, there are both non-negligible interactions and several important variables.

When the $a_j$ coefficients of the $G$ function are all equal and large, one is dealing with B-type function (B3, Equation 18, for which all $a_j$=6.42). By contrast, the case of $a_j$ null coefficients (1984) would correspond to a C-type function (e.g. functions C1 above).

## 2.1 Computational arrangements

Each test comparison is repeated 50 times in order to check the reproducibility of results, and to reduce the stochastic variation in the results to an acceptable level. Each repetition makes use of an equal number $N$ of quasi-random rows from the Sobol matrix. The total cost of the analysis is kept constant across methods. For each of the 50 repetitions, the randomization procedure (columns permutation) is based on the randomly assigning input factors to the various columns of the QMC matrices, rather than selecting new rows.

As customarily in QMC computations using Sobol' sequences the column dimension $N$ of each matrix is rounded to the nearest power of two: each power of two corresponds to a 'full scan' of the hyperspace for these sequences. Stopping the sequence before would be as painting a room and stopping a stroke of brush before the end of the wall. Various values of $p$ ($p = 2, 3, ... 15$) have been tested for selected functions (class A,B and C) against $N_T$. Following Saltelli et al. (2010) simulation results are presented in terms of mean absolute error (MAE) versus the total cost of the analysis, where MAE is defined as:

$$MAE = \frac{1}{50}\sum_{r=1}^{50} \left(\frac{\sum_{j=1}^{k}|\hat{T}_j - T_j|}{k}\right)_r \quad (22)$$

I.e. the total error over all factors is considered for the difference between the numerical estimate of the index (averaged over all available elementary effects) and the analytic value. Results are plotted on a decimal logarithmic scale.

# 3 Results
## 3.1 Testing the estimator of Glen and Isaacs (2012)
In Figure 2 some results are provided comparing the correlation-based estimators suggested by Glen and Isaacs (2012 – there named estimator D3) against Saltelli et al. (2010) for the six-dimensional $G$ function (Equation 12).

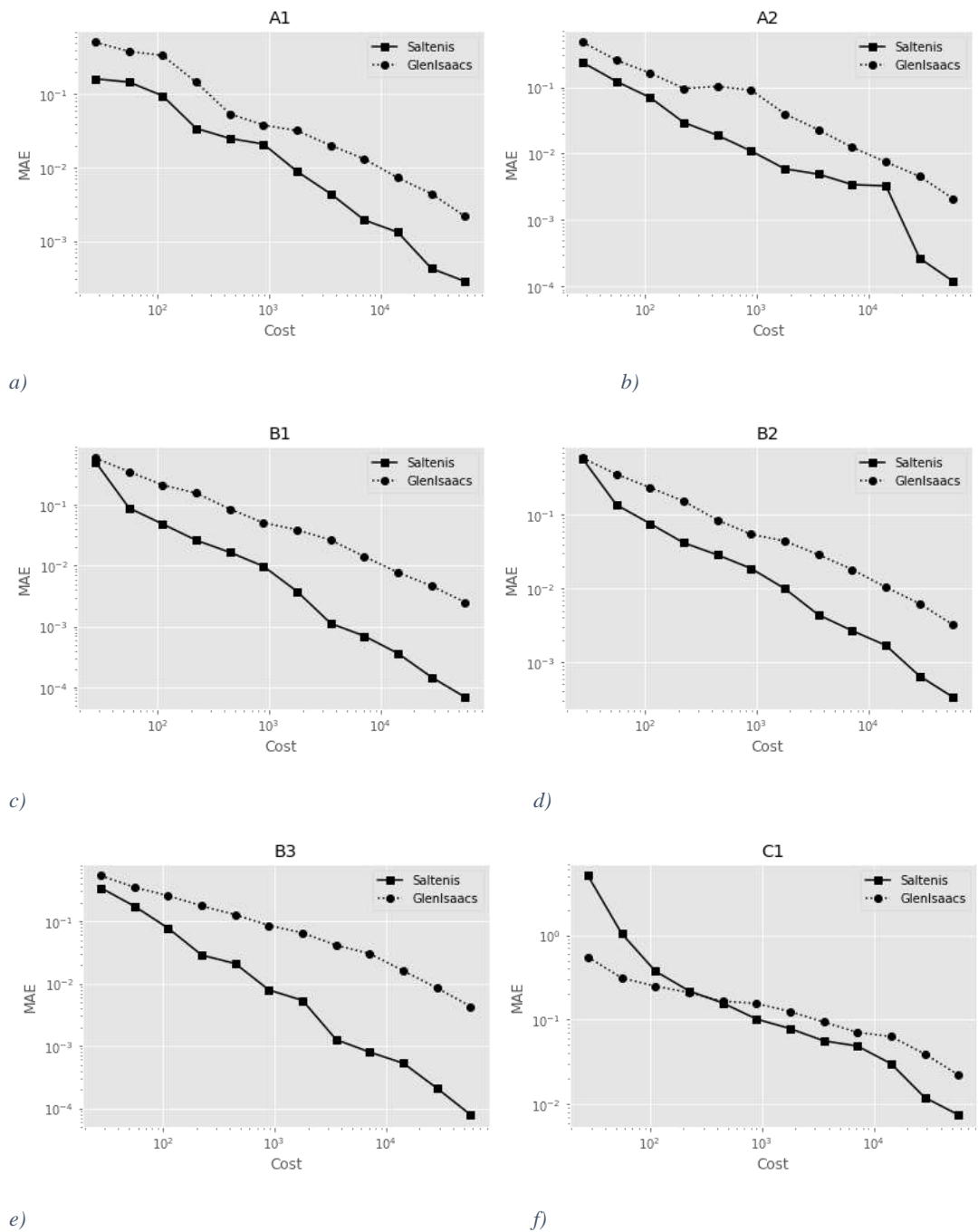

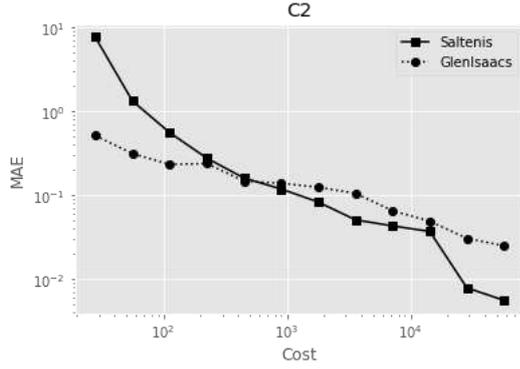

*g)*

*Figure 2(a,g) MAE vs cost ($N_T$) on logarithmic scale for the Šaltenis asymmetric estimator (square, continuous line) vs. the Glen-Isaacs estimator (circle, dotted line). Functions: A1, A2, B1, B2, B3, C1, C2 (Eq. 15-21). Python implementation.*

These results have been also confirmed by the Matlab® implementation for the function A2 (not shown here). As discussed above, computing $S_j$ requires couples of points where all factors but $X_j$ have differing values. The logic of correcting the correlation between these sets of points for spurious correlations is that when computing the correlation $\rho_j$ for the sensitivity index $S_j$ of factor $j$ we are considering e.g. vectors such as $f(\boldsymbol{a}_i)$ and $f\left(\boldsymbol{b}_{\boldsymbol{ai}}^{(j)}\right)$ with $i = 1,2,...N$, where the column $j$ is identical in the two vectors, and all other columns are different. If - because of the finite value of $N$ - any of these columns – say column $q$ – is spuriously correlated in the two matrices, then this spurious correlation should be removed from $\rho_j$ as described in Glen and Isaacs (2012). These authors suggest that a similar correction is useful for computing $T_j$. This claim is only partly supported by Figure 2(a-g), which shows how the estimator of Saltelli et al. (2010) for $T_j$ has better convergence properties at smaller sample size for functions of type A and B. This can be easily explained considering that the computation of $T_j$ requires vectors such as $f(\boldsymbol{a}_i)$ and $f\left(\boldsymbol{a}_{\boldsymbol{bi}}^{(j)}\right)$ with $i = 1,2,...N$, where now all columns but $j$ are identical in the two vectors, and the differing column $j$ is the one under investigation. There are no chances of spurious correlation in this case. The estimator of Šaltenis used in Saltelli (2010) and Sobol' (2001) presents a quicker convergence than the one developed by Glen and Isaacs.

We now focus on the design, comparing the case of $n = 2$ for both the symmetric and asymmetric case, as well as cases where $n > 2$ for the symmetric case. As already mentioned the case $n > 2$ for the asymmetric case was already tested in Saltelli et al., 2010, and led to poorer results when compared to the case $n = 2$.

### 3.2 Testing the optimal number of sample matrices

We take as starting point for the analysis the overall cost of the sensitivity analysis in terms of total numbers of runs $N_T$, and the number of factors $k$. Both $N_T$ and $k$ are in fact known to the modeller prior to the analysis. Based on these two parameters different couples of $n$ and $N$ can be chosen which meet the target $N_T$ value.

In terms of total number of effects which can be derived from the design, given for the symmetric case by $E_T = \frac{1}{2}Nkn^2(n-1)$, one would have interest to have $n$ as high as possible. In terms of discrepancy, one would instead wish to have lower values of $n$, which correspond to less repeated coordinates and hence in principle to a lower discrepancy. Note that the relation between discrepancy and error is not simple. A given discrepancy can be perfect for a smooth function and inadequate for a jigsaw-shaped one. Note also that when fixed $k$ and $N_T$ – e.g. in the formula $N_T = nN(1 + k(n-1))$ there is an inverse quadratic relationship between $n$ and $N$, describing the trade-off between economy and explorativity.

How to proceed? Let us illustrate our approach for a case where $k = 6$ and one can afford a cost $N_T \approx 500$ runs (Table 4). As $N$ needs to be a multiple of two in quasi-random number sequences based on Sobol' LP-TAU sequences the value of $N_T$ may deviate from 500. The discrepancy - estimated using the computational method provided by Jaeckel, 2002 and rounded to two digits – is also presented in Table 4.

**Table 4** - Possible values of $n$ and $N$ corresponding to model with $k = 6$ factors and an affordable total number of runs $N_T$ of about 500 and corresponding discrepancy D. $N_T$ has been adjusted as to have N as a multiple of 2, as requested in quasi Monte Carlo simulations based on Sobol' $LP_\tau$ sequences. The first row refers to the Saltelli et al., 2010 method where the total number of runs is $N_T = N(k + 1)$ and the total number of effects is $E_T = Nk$.

| $N$ | $n$ | $N_T$ | $E_T$ | $nN$ | $D$ |
|---|---|---|---|---|---|
| 64 | 2 (asymmetric) | 448 | 384 | 128 | 0.0065 |
| 32 | 2 (symmetric) | 448 | 384 | 64 | 0.0076 |
| 16 | 3 (symmetric) | 624 | 864 | 48 | 0.013 |
| 8 | 4 (symmetric) | 608 | 1152 | 32 | 0.020 |
| 4 | 5 (symmetric) | 500 | 1200 | 20 | 0.032 |
| 2 | 7 (symmetric) | 518 | 1764 | 14 | 0.053 |
| 1 | 10 (symmetric) | 550 | 2700 | 10 | 0.11 |

We also report in Table 4 the product $nN$, which – being the number of non-repeated coordinates- complements discrepancy as a measure of the explorativity of the design. The last row has the highest number of effects using the smallest number of random points – i.e. just ten rows of a single QMC matrix in six dimensions. The first row the opposite, as it uses as many as 64 rows of from two QMC matrices of 32 rows each but gives the smallest number of effect. In other words, the first row is the most explorative while the last one is the most economical in terms of effects per total number of runs.

Table 4 shows e.g. that in order to have discrepancy values of the order of 3% one should not exceed $n = 5$ for the case where $k = 6$ and $N_T \approx 500$ runs.

Table 4 tells us that one should not expect an improvement moving from the asymmetric to the symmetric case for $n = 2$, because we obtain the same number of effects at the cost of halving the explorativity of the design. This is confirmed in Figure 3 where we see that the asymmetric case of Saltelli (2010) systematically outperforms the symmetric one. Always based on the example of Table 4 one would have hoped that moving to $n > 2$ could improve the economy because the decrease in explorativity is in this case contrasted by an increased number of effects. This expectation is not borne by the results (Figure 3). The asymmetric design based on just two matrices is still the best design option to compute the total sensitivity indices $T_j$. It would thus appear that explorativity beats economy, e.g. that the increased number of effects does not overcome the decreased number of original coordinates.

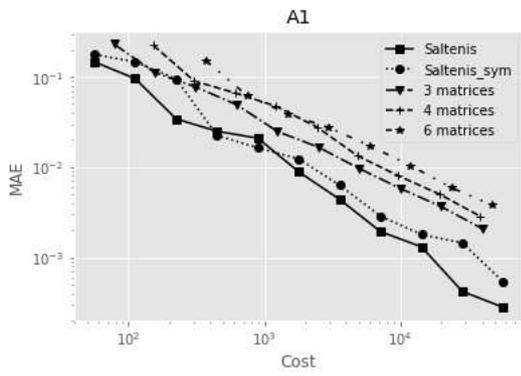
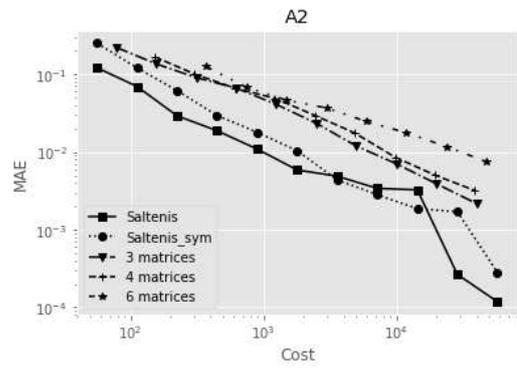

*a)* *b)*

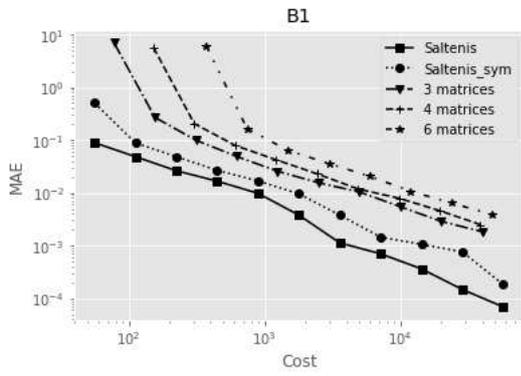
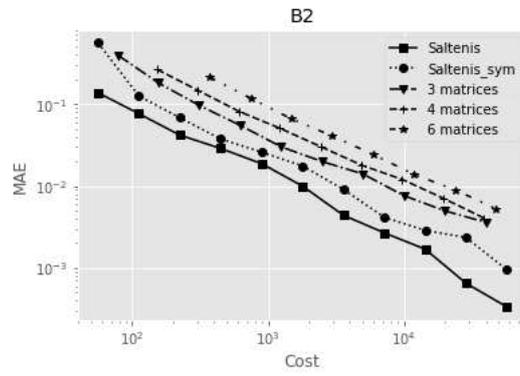

*c)* *d)*

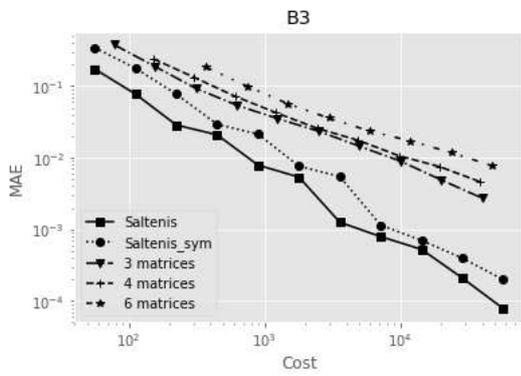
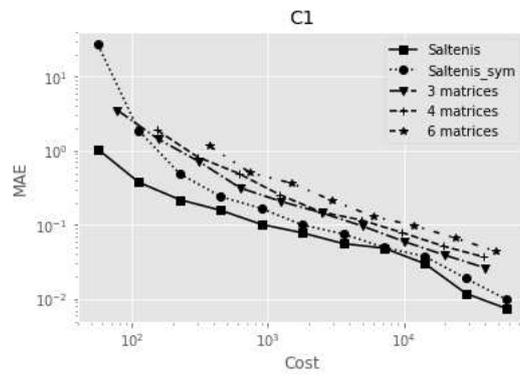

*e)* *f)*

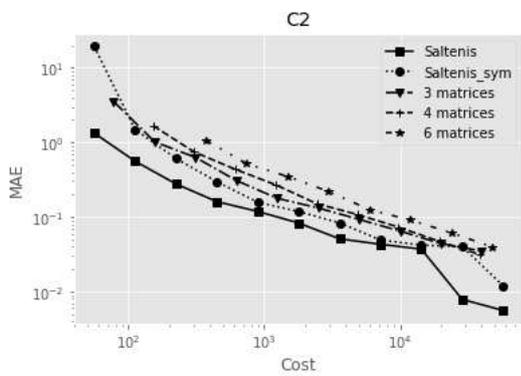

*g)*

*Figure 3 - MAE vs cost ($N_T$) on a logarithmic scale for the Šaltenis two-matrix-based asymmetric estimator (square, continuous line), two-matrix-based symmetric estimator (circle, dotted line), three-matrix-based estimator (triangle, dashed-dotted line), four-matrix-based estimator (cross, dashed line) vs. six-matrix-based estimator (star, dash-dot-dotted line). Functions: A1, A2, B1, B2, B3, C1, C2 (Eq. 15-21).*

The differences are also in this case modest and decreasing from A- or B-functions towards C-functions.

Additionally, Lamboni (2018) (equation 22) and Saltelli et al. (2010) estimators are compared for a variable number of matrices in Table 5. See Table 1 for a complete description of the indices reported in the equation.

$$E_{X_{\sim j}}\left(V_{X_j}(Y|X_{\sim j})\right) = \frac{n}{N(n-1)^2} \sum_{i=1}^{N} \sum_{l=a}^{n} \left[\sum_{m=a, m \neq q}^{n} \frac{1}{n-1} \left[f(\mathbf{h}_i) - f\left(\mathbf{h}_{mq}^{(j)}\right)\right]\right]^2 \quad (23)$$

The comparison refers to the symmetric Šaltenis estimator for the sake of consistency.

*Table 5 - Percentage differences on the MAE of Lamboni estimator against the estimator of Šaltenis for the different functions as well as number of matrices against the sample size N inquired.*

| Function | Number of matrices | 4 | 8 | 16 | 32 | 64 | 128 | 256 | 512 | 1024 | 2048 | 4096 | 8192 |
|---|---|---|---|---|---|---|---|---|---|---|---|---|---|
| A1 | 2 | 0 | 0 | 0 | 0 | 0 | 0 | 0 | 0 | 0 | 0 | 0 | 0 |
|    | 3 | 0 | 0 | 1 | 1 | 0 | -2 | -1 | 7 | 0 | -7 | -11 | -5 |
|    | 4 | -2 | 0 | 0 | 1 | 0 | -3 | -7 | -9 | -3 | -2 | 0 | 6 |
|    | 6 | 0 | -3 | -2 | -2 | -1 | -7 | -10 | -2 | 13 | 16 | 24 | 29 |
| A2 | 2 | 0 | 0 | 0 | 0 | 0 | 0 | 0 | 0 | 0 | 0 | 0 | 0 |
|    | 3 | 0 | -1 | 0 | 0 | 0 | 0 | -2 | -2 | -2 | -4 | -9 | -29 |
|    | 4 | -1 | -2 | 0 | 0 | 0 | 0 | 0 | 0 | 0 | -3 | -9 | -22 |
|    | 6 | 6 | 0 | -1 | 1 | 2 | 2 | 0 | 0 | -2 | -3 | -4 | -21 |
| B1 | 2 | 0 | 0 | 0 | 0 | 0 | 0 | 0 | 0 | 0 | 0 | 0 | 0 |
|    | 3 | 0 | 0 | 0 | 0 | 0 | -1 | -1 | -1 | 0 | -1 | 0 | 0 |
|    | 4 | 0 | 0 | 0 | 0 | -1 | -1 | -1 | 0 | 0 | -1 | -3 | -9 |
|    | 6 | 0 | 0 | 0 | -1 | 0 | 0 | 0 | 0 | 0 | 0 | 2 | 6 |
| B2 | 2 | 0 | 0 | 0 | 0 | 0 | 0 | 0 | 0 | 0 | 0 | 0 | 0 |
|    | 3 | 0 | -1 | 0 | -1 | -2 | 0 | -3 | -4 | -1 | -6 | -7 | -9 |
|    | 4 | 0 | -1 | -1 | 0 | -1 | -2 | -4 | -8 | -3 | -5 | -14 | -20 |
|    | 6 | -1 | -3 | -3 | -2 | -6 | -5 | -10 | -10 | -7 | -6 | -1 | 0 |
| B3 | 2 | 0 | 0 | 0 | 0 | 0 | 0 | 0 | 0 | 0 | 0 | 0 | 0 |
|    | 3 | 0 | 0 | 0 | 0 | 0 | 0 | 0 | 0 | 0 | 0 | 0 | 0 |
|    | 4 | 0 | 0 | 0 | 0 | 0 | 0 | 0 | 0 | 0 | 0 | 0 | 0 |
|    | 6 | 0 | 0 | 0 | 0 | 0 | 0 | 0 | 0 | 0 | 0 | 0 | 0 |
| C1 | 2 | 0 | 0 | 0 | 0 | 0 | 0 | 0 | 0 | 0 | 0 | 0 | 0 |
|    | 3 | 0 | 1 | 0 | -1 | -2 | -2 | -3 | -2 | -1 | 0 | 2 | -2 |
|    | 4 | 0 | -3 | 3 | -3 | -3 | -3 | -4 | -1 | -2 | -1 | -10 | -17 |
|    | 6 | 0 | -2 | -1 | 0 | 0 | -2 | -5 | 0 | -1 | -4 | -9 | -13 |
| C2 | 2 | 0 | 0 | 0 | 0 | 0 | 0 | 0 | 0 | 0 | 0 | 0 | 0 |
|    | 3 | -3 | 12 | -5 | -1 | 0 | 1 | 0 | 1 | 0 | 0 | -1 | -2 |
|    | 4 | -5 | -7 | -1 | 0 | 0 | 2 | 3 | 0 | 0 | -1 | -1 | -2 |
|    | 6 | -10 | -4 | 1 | 1 | -1 | 0 | -1 | 0 | 1 | 2 | 4 | 2 |

The difference in Table 5 is defined as (MAE$_{\text{šaltenis}}$ − MAE$_{\text{Lamboni}}$)/MAE $_{\text{šaltenis}}$. The comparison has been performed for a set of 2-6 symmetric matrices cross cutting the pool of A-, B- and C-type functions. Negative differences point at a quicker convergence of the šaltenis estimator (MAE$_{\text{šaltenis}}$ < MAE$_{\text{Lamboni}}$), the reverse is true for positive difference. The column label is the column dimension $N$. Lamboni is outperformed by the šaltenis estimator and the gap in performance in comparison with the asymmetric design is even higher (not reported). This result can be attributed to the lower explorativity of Lamboni's design. As one can notice from the formula used by Lamboni (Eq. 22), the effects of scrambling columns from mixed matrices (e.g. $f(a_c^j)$ against $f(a_b^j)$) is not considered in this type of estimator, which only compares the effect of matrices with scrambled columns against sample matrices. I.e., this design resorts to a lower explorativity at the same cost.

In conclusion we cannot reproduce the results of Lamboni (2018).

### 3.3 Improving the existing best practices: a dynamical sampling procedure

As shown above, the trade-off between economy and exploration demonstrates that n>2 is not a convenient design choice. In the following we tested a number of dynamical sampling procedure to improve the estimation method of Saltelli et al. (2002).

The idea consists of investing the computational time more efficiently by improving, model execution after model execution, the $T_j$'s estimation for the subset of the most important factors, while devoting less computational resources to the least important ones. To achieve this, we first estimate all $T_j'$ s for an arbitrarily fixed number of run $N_{Ts}$ (where $Ts$ stands for threshold) so that all $T_j'$ s estimates have achieved a modicum of stability. For the subsequent runs we then select the subset of factors to be resampled using a variety of methods, including

- A Russian roulette approach, where each factor has a probability of being selected that is proportional to its relative $T_j$ as known at that point in the simulation. In this way the more a factor is important (and the larger hence its expected contribution to the MAE), the more estimations it receives.

- Simply comparing at $N_{Ts}$ the standard deviations of the numerator for the formula for $T_j$ in order to restrict subsequent calculations to the factors with the largest standard deviations. This could be done either
    - By redistributing the computation of the effects within an existing point, or
    - By generating new points where just the important factors are resampled.

For the purpose of illustration, we put it in numbers in Box 1 to show how it plays.

# BOX 1. More effects for the important factors.

After warming the system with a fraction $f_1 = N_{TS} / N_T$ of the available total number of calls to the function whose $T_i$'s are being estimated, a fraction $f_2$ of the input factors considered as least important are left 'as they are', meaning that their estimated index $\widehat{T_i}$ is frozen at the value known at $N_{TS}$. To illustrate the procedure let us take $f_1 = \frac{1}{4}, f_2 = \frac{1}{3}$, i.e. the system is warmed using one fourth of the available runs at which point one third of the factors, those with the smallest $\widehat{T_j}$ are fixed. The remaining runs after warming, i.e. $N_T - N_{TS}$, are now divided among the $k - k/3 = 2/3 \, k$ factors.

The table below show what happens for a system with $k = 6$ factors when fixing 2 factors for values of the base sample matrix varying between $2^4$ and $2^7$ and total sample size varying between 112 and 896. The first base sample of 16 rows tells us that fixing two factors implies that each of the remaining four factors will have 22 (4+12+6) estimates instead of 16, with a gain of about one third. This effect remains constant over the rows, i.e. for a base sample of 128 one gets 176 (32+96+48) runs per factor instead of 128.

| $N$ | $N_T = N(k+1)$ when $k=6$. Each factor receives $N_b$ estimates | Warm up $N_{TS} = \frac{N}{4}(k+1)$ (number of base runs used) | $N_T - N_{TS}$ and (number of base runs left) | Runs to estimate the remaining $2/3 \, k = 4$ factors | Runs saved | Extra runs per factor. It is easy to say that these are one fourth of the runs saved |
|---|---|---|---|---|---|---|
| 16 | 16*7=112 | 28 (4) | 84 (12) | 12*(4+1)=60 | 84-60=24 | 24/4=6 |
| 32 | 224 | 56 (8) | 168 (24) | 120 | 48 | 12 |
| 64 | 448 | 112 (16) | 336 (48) | 240 | 96 | 24 |
| 128 | 896 | 224 (32) | 672 (96) | 480 | 192 | 48 |

How do we deploy these extra runs per factor? Taking the first base sample of 16 as example, for each of the four selected factors six new effects should be placed in six different existing points, chosen among the 12 available, in such a way that no point receives more than an extra effect. Note that this allows in each point the computation of two new effects for the given factor as there will now be three estimates for a given factor along the same axis. I.e., the original base point from matrix A, the corresponding point from matrix AB, and the extra point X, give rise to the effects A-AB, A-AX, AB-AX.

Alternatively, still for a base sample of 16, one can spend the 24 extra runs to create new points, extending the base sample matrix. If for each of the new points, five runs are needed (one for the A matrix plus four for the factors estimated), one can add a maximum of five new stars using 25 runs. One of these runs can be eliminated afterwards as to maintain the cost constant across approaches. This yields just five extra effects per factor for three factors and four for one, instead of six. Hence, the number of effects computed would be 20, i.e. 4+12+4, or 21, i.e. 4+12+5 for a base sample of 16 and a number included in the range 160-168, i.e. 32+96+32/40 for a base sample of 128 dependent on the runs taken out from some of the extra stars to balance out the cost.

The cases of a sample base consisting of 32, 64 and 128 rows detailed in Box 1 above are demonstrated in Figure 4 below. The two approaches are compared against the standard Šaltenis estimator. We examine in this case the trend at a smaller sample size as to evaluate the potential of the new method when the number of run available is limited. The results of the computation are plotted out on every row rather than entire block unlike the previous charts as to appreciate the trend row-wise.

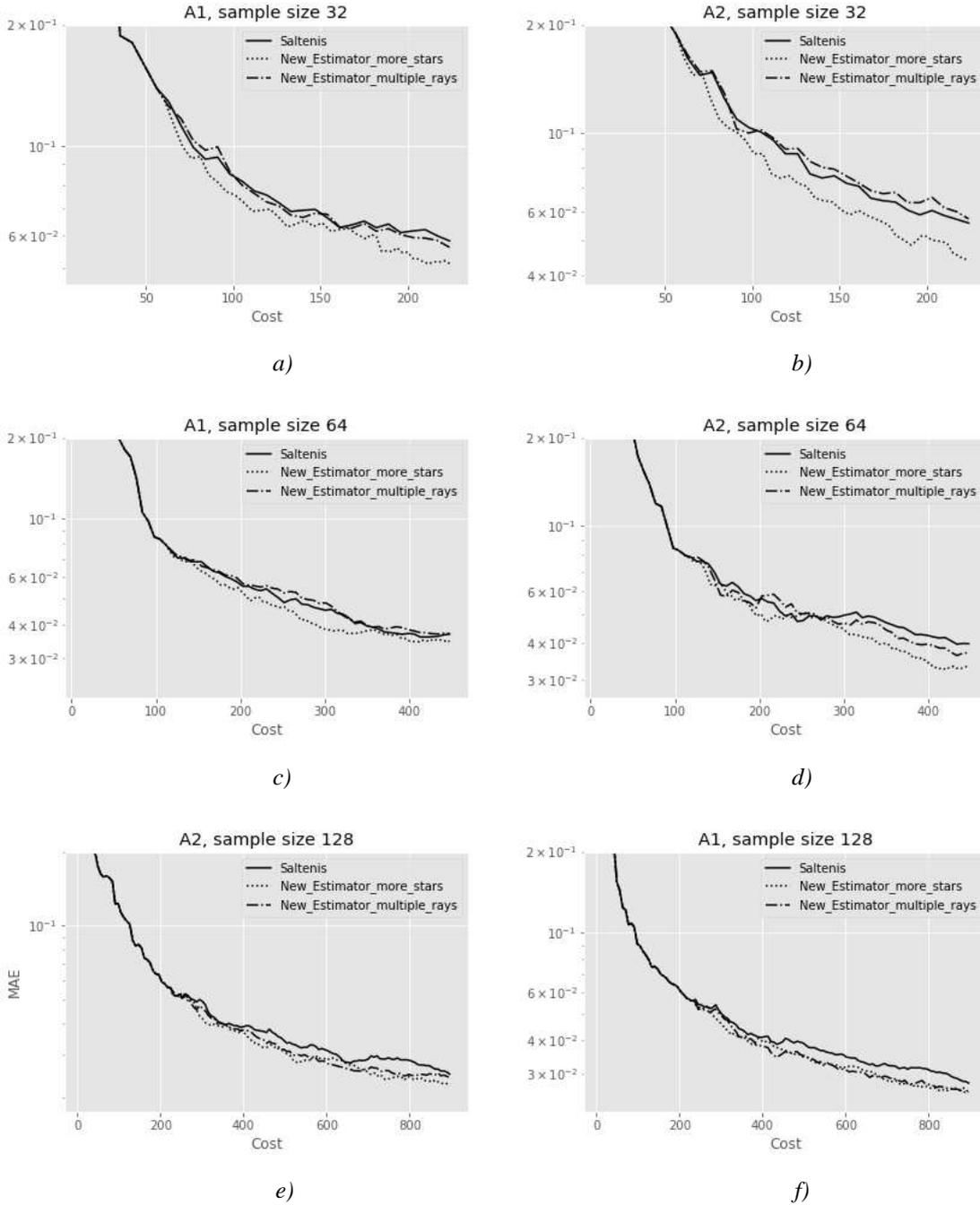

*Figure 4 - MAE vs cost ($N_T$) on a y- semi-logarithmic scale for the Šaltenis two-matrix-based asymmetric estimator (continuous line), and the new multiple-ray-based estimator (dash-dotted line), and the extra-star-based estimator (dotted line).. Functions: A1 and A2 (Eq. 15-16).*

In general, the approach adding extra stars with less rays appears to perform better, although the improvement is limited and dependent on the function. In fact, function A2 shows in general more tangible improvements over A1, although these effects do not scale up with the sample size. The approach of adding extra rays to the same star is affected by a higher variability: the improvements are scant or it is even performing worse than the Šaltenis estimator under particular arrangements.

None of the approaches attempted gave fully satisfying results, in that improvements were limited, function dependent, and relying on ad hoc design parameters such as $f_1$ and $f_2$ in Box 1. One problem worth mentioning is that even allowing for large $N_{TS}$ still the measures may be oscillating, with the results of excluding factors which are instead relevant.

Additionally, when the new effects are computed on the existing points the efficiency increase, though not the explorativity. When instead new points are generated, the explorativity is increased, while the efficiency deteriorates: after eliminating from the computation $q$ factor the efficiency reduces to $\frac{k-q}{1+k-q}$ which is smaller than $\frac{k}{1+k}$. E.g. eliminating 2 factors in the 6 factors function used in this work moves the efficiency from 6/7 (.86) to 4/5 (.80). Eliminating 4 factors would move it even lower to 2/3 (.67). However, our findings demonstrate that a less efficient but more explorative design is to be preferred over the opposite arrangement.

# 4 Conclusions

Taking the works of Saltelli et al., 2010, Glen and Isaacs (2012) and Lamboni (2018) as our point of departure we have explored different estimators and different number of sample matrices to improve the computation of the total effect index $T_j$ using a taxonomy of test functions due to Kucherenko et al., 2011).

We see that the estimator of Glen and Isaacs (2012) is outperformed by Šaltenis (1982) although results appear to depend upon the test function inquired.

We failed to improve computational results by extending the symmetric matrix arrangement to values of $n > 2$ due to the fact that the larger number of effects obtained with $n > 2$ is not enough to compensate for the loss of explorativity, also evidenced by our discrepancy calculation. This result is in disagreement with Lamboni (2018).

Finally, we sought to achieve an improvement against the existing best practice by distributing the available number of simulations in such a way as to estimate factors' effect according to their importance. This approach yielded improvements, but of a limited nature. The quest for a better sample-based estimator for $T_j$ remains open.

# 5 Acknowledgments

Sergei Kucherenko from Imperial College of London and Thierry Mara from the Joint Research Centre of the European Commission offered useful comments and suggestions. The remaining errors are uniquely due to the authors. This work was partially funded by a Peder Sather grant of the university of Berkeley "Mainstreaming Sensitivity Analysis and Uncertainty Auditing", awarded in June 2017. Arnald Puy has worked on this paper on a Marie Sklodowska-Curie Global Fellowship, grant number 792178.

# Appendix 1 - Repeated coordinates

Imagine that an arbitrary number $N_T$ is to be used in a hypercube in $k$ dimensions. This would need – bar random overlaps – a total of $N_T k$ coordinates, and an equal number of random numbers.

Any calculation of the total sensitivity indices $T_j$ constrains and reduces this total number for the need of having point which differ only for the value of one coordinate. There are different arrangements to do this, and they can be compared by how many coordinates they use out of the maximum number of $N_T k$ coordinates. The lower this fraction the less explorative the design will be, though this is done to the effect of increasing the design's economy.

**Couples**. The $N_T$ points can be arranged in $N_T/2$ couples. Each couple differ for one or another of the $k$ coordinates. To produce this arrangement, one need to generate $N_T(k+1)/2$ coordinates (of e.g. random numbers). Each couple needs just $k$ coordinates for one point and only one extra coordinate for the companion point. Thus, the fraction $f$ of coordinates relative to the above-mentioned maximum of $N_T k$ coordinates is

$$f = \frac{N_T(k+1)/2}{N_T k} = \frac{k+1}{2k} \tag{A1}$$

Which tends to ½ for increasing $k$.

**Stars.** A star arrangement is one where $\frac{N_T}{k+1}$ points are generated in the hypercube, and from each of this point each of the available *k* dimensions is explored in turn. It this way each star is made of *k+1* points, the total number of points is still *N_T*, and each star needs *k+k* coordinates, *k* for the centre point of the star and one for each of its *k* rays. Thus, the fraction *f* is now:

$$f = \frac{2N_T k}{N_T(k+1)k} = \frac{2}{k+1} \tag{A2}$$

and decreases with increasing k.

**Winding stairs (one trajectory).** In a winding-stair design the hypercube is explored by a curve whereby each coordinate is increased in turn. This design needs *k* coordinates to generate the first point and *N_T-1* additional coordinates for the remaining points. This one gets

$$f = \frac{N_T + k - 1}{N_T k} \tag{A3}$$

Which tends to *1/k* as generally *N_T>>k*.

**Saltelli 2001, asymmetric.** We now introduce a new symbol *N* to denote the column dimension of a generic matrix. This number will not appear in the results for the fraction *f*. The design needs one base matrix *A* and *k* additional matrices $A_B^1, A_B^2, \ldots A_B^k$. This corresponds to a total of *N(k+1)* points which in principle would correspond to a total of *Nk(k+1)* coordinates, but in facts only needs a total of *2Nk* coordinates, *Nk* for each of the two matrices *A* and *B*. Hence

$$f = \frac{2Nk}{Nk(k+1)} = \frac{2}{(k+1)} \tag{A4}$$

Same as 'star' above.

**Saltelli 2001, symmetric.** In this design one make use of both sets $A_B^1, A_B^2, \ldots A_B^k$ and $B_A^1, B_A^2, \ldots B_A^k$, for a total of *2N(k+1)* points which in principle would correspond to a total of *2Nk(k+1)* coordinates, which instead still needs only *2Nk* coordinates, *Nk* for each of the two matrices *A* and *B*. Hence

$$f = \frac{2Nk}{2Nk(k+1)} = \frac{1}{(k+1)} \tag{A5}$$

**Generalization of the symmetric case.** The design now includes *n* base matrices *A,B,…X* where *X* is the n[th] matrix, plus a total of twice $\binom{n}{2} k$ additional hybrid matrices of the type $A_B^1, A_B^2, \ldots A_B^k, B_A^1, B_A^2, \ldots B_A^k, \ldots$ where the binomial corresponds to the possible number of coupling of two matrices. Thus, the total number of matrices is $n + 2\binom{n}{2} k = n + n(n-1)k = n(1 + k(n-1))$. The total number of points is hence $nN(1 + k(n-1))$, corresponding to a maximum number of coordinates $nNk(1 + k(n-1))$. Since the number of coordinates used in this design is just those of the base matrices, i.e. *nNk*, the fraction *f* is

$$f = \frac{nNk}{nNk(1+k(n-1))} = \frac{1}{1+k(n-1)} \tag{A6}$$

Which decreases with increasing *n* and reduces to (A5) for *n=2*.